\documentclass{article}

\usepackage{microtype}
\usepackage{graphicx}
\usepackage{subcaption}
\usepackage{booktabs} 

\usepackage{silence}
\usepackage{hyperref}
\usepackage{xurl}
\usepackage{fvextra}
\usepackage{multirow}
\usepackage[dvipsnames,table]{xcolor}

\usepackage{listings}
\usepackage{xspace}
\usepackage{enumitem}
\usepackage[flushleft]{threeparttable}
\usepackage{makecell}
\usepackage{tikz}
\usepackage{wrapfig}
\usepackage{textcomp}  
\usepackage{scalerel}  
\usepackage{tipa}
\usepackage{pifont}
\usepackage[subtle]{savetrees}
\usepackage[most]{tcolorbox}
\usepackage[normalem]{ulem}

\newcommand{\tech}{\textsc{GCC}\xspace}

\newcommand{\autocoderover}{AutoCodeRover\xspace}
\newcommand{\aider}{Aider\xspace}
\newcommand{\sweagent}{SWE-agent\xspace}
\newcommand{\coder}{CodeR\xspace}
\newcommand{\ibmagent}{IBM Research Agent-101\xspace}
\newcommand{\opencsgagent}{OpenCSG StarShip\xspace}
\newcommand{\bytedanceagent}{Bytedance MarsCode\xspace}
\newcommand{\amazonqagent}{Amazon Q Developer\xspace}

\newcommand{\repounderstander}{RepoUnderstander\xspace}
\newcommand{\lingma}{Alibaba Lingma Agent\xspace}
\newcommand{\factorydroid}{Factory Code Droid\xspace}
\newcommand{\opendevincodeact}{OpenDevin+CodeAct v1.8\xspace}
\newcommand{\codestoryaide}{CodeStory Aide\xspace}
\newcommand{\mentatbot}{MentatBot\xspace}
\newcommand{\honeycomb}{Honeycomb\xspace}
\newcommand{\gru}{Gru\xspace}
\newcommand{\isoform}{Isoform\xspace}
\newcommand{\supercoder}{SuperCoder2.0\xspace}
\newcommand{\repograph}{RepoGraph\xspace}  %
\newcommand{\moatless}{Moatless\xspace}
\newcommand{\rag}{RAG\xspace}
\newcommand{\specrover}{SpecRover\xspace}
\newcommand{\masai}{MASAI\xspace}
\newcommand{\sima}{SIMA\xspace}
\newcommand{\appmapnavie}{AppMap Navie\xspace}

\newcommand{\gptfour}{GPT-4\xspace}
\newcommand{\gptfouro}{GPT-4o\xspace}

\newcommand{\claude}{Claude\xspace}

\newcommand{\claudesonnet}{Claude 3.5 S\xspace}
\newcommand{\claudeopus}{Claude 3 Opus\xspace}

\newcommand{\chatgpt}{GPT-3.5\xspace}

\newcommand{\llm}{LLM\xspace}

\newcommand{\Comment}[1]{}

\definecolor{yucky}{HTML}{a64d79}

\definecolor{codegreen}{rgb}{0,0.6,0}
\lstdefinestyle{mystyle}{  
    commentstyle=\color{codegreen},
    keywordstyle=\color{blue},
    basicstyle=\ttfamily\small,
    breakatwhitespace=false,        
    breaklines=true,                 
    captionpos=b,                    
    keepspaces=true,                 
    showspaces=false,                
    showstringspaces=false,
    showtabs=false,                  
    tabsize=2
}
\usepackage[preprint]{icml2026}

\usepackage{amsmath}
\usepackage{amssymb}
\usepackage{mathtools}
\usepackage{amsthm}

\usepackage[capitalize,noabbrev]{cleveref}

\theoremstyle{plain}

\theoremstyle{definition}

\theoremstyle{remark}

\usepackage[textsize=tiny]{todonotes}

\icmltitlerunning{Git Context Controller: Manage the Context of Agents by Agentic Git}

\begin{document}

\twocolumn[
  \icmltitle{Git Context Controller: Manage the Context of Agents by Agentic Git}




  \begin{icmlauthorlist}
    \icmlauthor{Junde Wu}{ox}
    \icmlauthor{Minhao Hu}{ox}
    \icmlauthor{Jiayuan Zhu}{ox}
    \icmlauthor{Jiazhen Pan}{tum}
    \icmlauthor{Yuyuan Liu}{ox}
    \icmlauthor{Min Xu}{cmu}
    \icmlauthor{Yueming Jin}{nus}
  \end{icmlauthorlist}

  \icmlaffiliation{ox}{University of Oxford}
  \icmlaffiliation{tum}{Technical University of Munich}
  \icmlaffiliation{cmu}{Carnegie Mellon University}
  \icmlaffiliation{nus}{National University of Singapore}

  \icmlcorrespondingauthor{Junde Wu}{jundewu@ieee.org}
  \icmlcorrespondingauthor{Yueming Jin}{ymjin@nus.edu.sg}

  \vskip 0.3in
]



\printAffiliationsAndNotice{}  

\begin{abstract}
Large language model (LLM) agents have demonstrated strong capabilities in long-horizon tasks by interleaving reasoning with tool use. However, as these agents scale to complex workflows such as software engineering and open-ended research, context management becomes a fundamental bottleneck: interaction histories grow unbounded, become costly to maintain, and are difficult to reuse across sessions and agents.
We introduce \textbf{Git-Context-Controller (GCC)}, a structured context management framework inspired by software version control systems. GCC elevates agent context from a transient token stream to a persistent, navigable memory workspace with explicit operations---\texttt{COMMIT}, \texttt{BRANCH}, \texttt{MERGE}, and \texttt{CONTEXT}, that enable milestone-based checkpointing, isolated exploration of alternative reasoning paths, and hierarchical retrieval of historical context. By organizing agent memory as a versioned file system, GCC allows agents to manage long-term goals, recover and transfer reasoning across sessions, and coordinate multi-trajectory problem solving in a principled manner.
Empirically, agents equipped with GCC achieve state-of-the-art performance on both SWE-Bench and BrowseComp benchmarks. On SWE-Bench Verified, GCC improves task resolution by over 13\% relative to strong long-context baselines and outperforms 26 existing open and commercial systems, reaching over 80\% success rate. https://github.com/ImprintLab/git-context-controller.
\end{abstract}

\section{Introduction}
LLM-based agents have been capable of interleaving internal chain-of-thought reasoning with external tool calls \cite{wu2025agentic}. Such architecture has shown strong performance in decision-making tasks, web interaction, and question answering benchmarks, providing a foundation for more sophisticated agents. In software engineering domains, frameworks like SWE-Agent \cite{sweagent} used similar paradigm by integrating code generation, execution, and test loops to implement iterative software development (e.g., writing, compiling, debugging). Following this idea, production-grade tools such as Anthropic’s Claude Code and Google’s Gemini CLI bring LLM-based agents to the command line, enabling code completion, debugging, and search within a single session. 

However, as LLM agents are increasingly deployed for long-horizon reasoning in complex, large-scale workflows, context management emerges as a fundamental bottleneck. A common issue observed in coding agents is that sessions gradually forget previous context and become increasingly costly as the context grows longer. Starting a new session typically erases the agent’s memory of prior goals, user preferences, and task-specific instructions. As a result, users are forced to repeatedly provide the same context in every new session.
Current solutions rely on several common strategies. A straightforward approach is to truncate older context once the token limit is reached. This risks discarding important historical details—especially problematic when the agent needs to revisit earlier decisions or maintain consistency across multi-step plans. A more balanced approach compresses earlier reasoning into high-level summaries or to-do lists, as seen in systems such as Claude Code and Gemini CLI. These systems use summary-based anchors for future reasoning and persist abstracted task state (e.g., via \texttt{agent.md} file). However, relying on simple compression removes fine-grained details and weakens the agent’s ability to ground its actions in specific prior reasoning. A common and intuitive observation is that agents become ``dumber'' each time their context is compressed.
In conclusion, context is currently either \textit{too verbose} to be reusable or \textit{too abstract} to support concrete continuation and extension.

\begin{figure*}[t]
\centering
\includegraphics[width=0.95\linewidth]{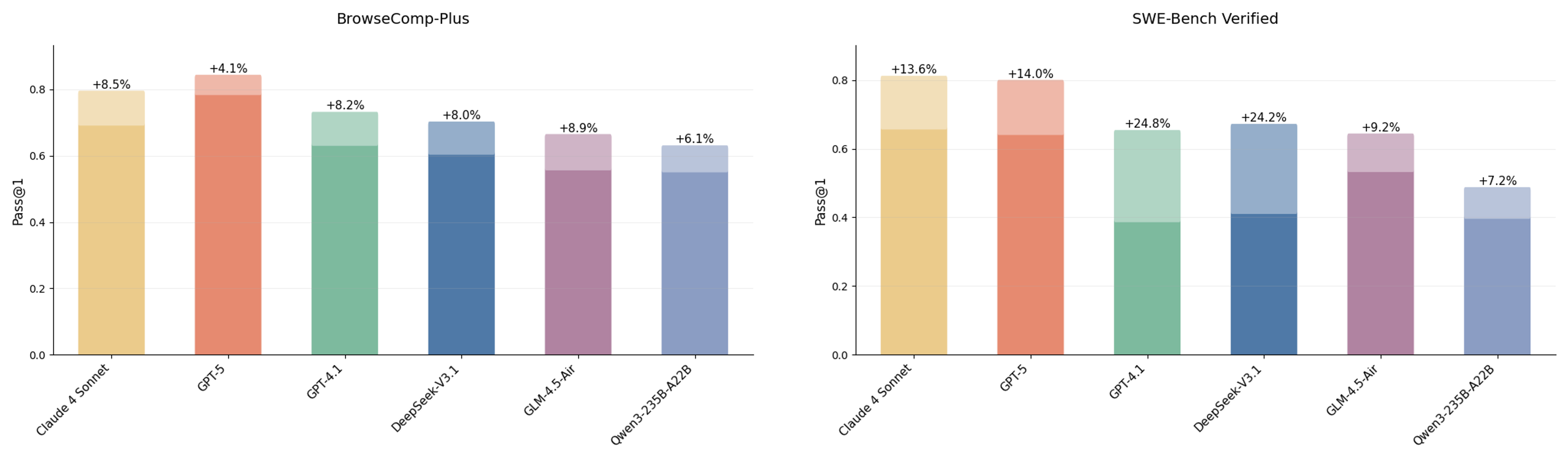}
\vspace{-5pt}
\caption{Results on BrowseComp-Plus and SWE-Bench Verified comparing baseline model performance with improvements achieved by equipping models with GCC.}
\label{fig:performance1}
\vspace{-10pt}
\end{figure*}

These limitations highlight the need for a more principled and structured approach to how AI agents log, manage, and retrieve context. Our key insight is that the challenges faced by long-horizon agents closely mirror those encountered by software engineers managing complex, evolving codebases. Inspired by the success of Git in software version control, we propose \textit{Git-Context-Controller (GCC)}, an agentic context control mechanism that elevates context management to an explicit abstraction layer. It organizes contextual information as a structured, version-controlled file system, and introduces a set of specialized commands designed to support logging, managing, and retrieving context across agentic workflows.

We implement this design through a standalone \texttt{Git-Context-Controller}, which structures agent context as a version-controlled file system under a unified \texttt{.GCC/} directory. Each project maintains a global roadmap (\texttt{main.md}), while each branch contains its own commit summaries, execution traces, and structured metadata. Agents interact with this controller through a small set of core commands: \texttt{COMMIT} to checkpoint meaningful progress, \texttt{BRANCH} to explore alternate strategies, \texttt{MERGE} to synthesize divergent reasoning paths, and \texttt{CONTEXT} to retrieve historical information at varying resolutions. These commands and data structures are provided to the agent and autonomously invoked to support long-horizon reasoning.

Such a design provides several complementary benefits. It enables multi-level context retrieval, allowing agents to access information at different levels of abstraction, ranging from high-level project plans to fine-grained OTA (Observation–Thought–Action) traces. Agents can flexibly navigate across these layers, starting from a coarse summary and drilling down into detailed execution histories whenever necessary, which makes past reasoning easy to trace and reuse. At the same time, the branching mechanism offers isolated workspaces for exploration, where agents can freely test new ideas or iterate without interfering with the main reasoning trajectory. This preserves focus while still allowing side explorations to be revisited or merged back into the primary workflow. Moreover, the framework naturally supports cross-agent and cross-session continuity: a new agent does not need to be re-instructed from scratch, and even an agent running on a different LLM or machine can seamlessly resume from the exact state left by its predecessor. This design facilitates smooth distribution and handover of agent-generated code and reasoning artifacts, in a manner analogous to how human developers collaborate through Git repositories.

Empirically, agents equipped with GCC achieve state-of-the-art (SOTA) performance on SWE-Bench and BrowseComp, two of the most widely used benchmarks for long-horizon coding and web-browsing reasoning. As shown in Fig.~\ref{fig:performance1}, on SWE-Bench, simply integrating GCC boosts models such as GPT-4.1 and DeepSeek by over $24\%$. Claude-4-Sonnet equipped with GCC further improves by $13.6\%$, reaching $80.2\%$, which constitutes the current state-of-the-art result on SWE-Bench. On BrowseComp-Plus, GPT-5 with GCC achieves $83.4\%$, also establishing a new state-of-the-art performance on this benchmark.

In summary, our contributions are:
\begin{itemize}
\item  We propose a novel view of agent memory as a dynamic, navigable codebase, complete with log files, branching histories, and metadata. This reframes context not just as passive history but as an evolving, queryable interface that supports both recall and structural reasoning.

\item We introduce GCC, a structured context management framework for LLM agents that integrates version control semantics—such as \texttt{COMMIT}, \texttt{BRANCH}, and \texttt{MERGE} into the reasoning loop. GCC organizes agent memory into persistent, interpretable artifacts that support long-horizon workflows, architectural modularity, and reproducibility.

\item When equipped with GCC, LLM-based agents achieve state-of-the-art empirical performance on both SWE-Bench and BrowseComp. In particular, on SWE-Bench, our method outperforms 26 existing systems (including both open-source and commercial models), achieving over 80\% task resolution rate.

\end{itemize}

\begin{figure*}[t]
\centering
\includegraphics[width=\linewidth]{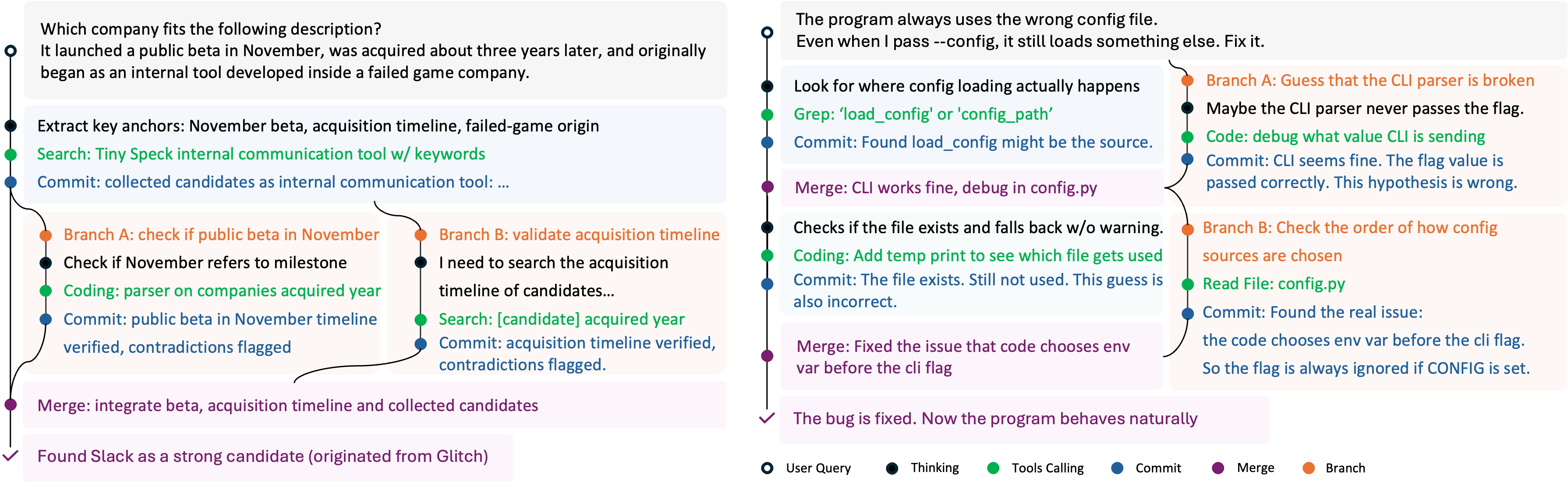}
\vspace{-5pt}
\caption{Illustration of GCC in action across two workflows: web-search reasoning and software debugging—showing how agents branch, explore, and merge structured context during long-horizon tasks.}
\label{fig:cover}
\vspace{-10pt}
\end{figure*}

\section{Method}
We proposed \texttt{Git-Context-Controller} (GCC) as an abstraction layer for agent memory, consisting of a structured file system paired with a series of callable commands that agents use to externalize, organize, and retrieve their reasoning, as shown in Fig. \ref{fig:cover}. Inspired by version control systems like Git, GCC transforms the agent’s ephemeral context into a persistent and navigable workspace. The core of GCC is a directory-based data structure that systematically organizes all historical context (GCC File System), together with a set of commands that enable agents to interact with and manipulate this structure (GCC Commands). Concretely, agents communicate with the controller via commands such as \texttt{COMMIT}, \texttt{BRANCH}, \texttt{MERGE}, and \texttt{CONTEXT}, which operate over a hierarchical workspace rooted at \texttt{.GCC/}. This workspace contains global planning artifacts (\texttt{main.md}), branch-specific execution traces (\texttt{log.md}), milestone-level summaries (\texttt{commit.md}), and structured metadata (\texttt{metadata.yaml}) that record architectural decisions and project states.
In the following, we provide a detailed introduction to the structured file system and commands of GCC.

\subsection{GCC File System}

GCC organizes agent context into a structured directory rooted at \texttt{.GCC/}, reflecting a three-tiered hierarchy of reasoning: high-level planning, commit-level summaries, and fine-grained execution traces. Each file in this hierarchy plays a distinct role in tracking the agent’s thoughts, progress, and architectural context. All files are plain-text and continuously updated through agent-invoked commands.

The overall layout of the file system is structured as following:
\begin{lstlisting}
.GCC/
|-- main.md           # a global roadmap summarizing project high-level intent, milestones, and shared planning state across all branches
|-- branches/
    |-- <branch-name>/
        |-- commit.md     # recording the progress of each commit 
        |-- log.md        # a detailed execution trace of OTA cycles, continuously recorded during the agent reasoning loop
        |-- metadata.yaml # a structured file storing branch-specific architectural and contextual metadata (file structures, dependencies, configs)
    |-- <another-branch>/...
\end{lstlisting}
In which, \textbf{\texttt{main.md}} sits at the root of the \texttt{.GCC/} directory and stores the global project roadmap. It records high-level project goals, key milestones, and the to-do list for development. This file is shared across all branches and serves as the canonical source of the project’s overall intent. The agent is prompted to initialize this file with the project goal and initial to-do list at the beginning of the project. It may be revised later when a conclusion is reached, a major outcome is completed, or significant changes to the roadmap occur. Such updates are optionally triggered after \texttt{COMMIT}, \texttt{MERGE}, or \texttt{BRANCH} by the agents.

Each branch has its own directory under \texttt{branches/}, which contains three primary files. The first is \textbf{\texttt{commit.md}}, a structured summary log that captures the evolving progress of the branch. Each time the agent calls \texttt{COMMIT}, the controller appends a new entry to \texttt{commit.md} following a standardized template consisting of three blocks: (1) \textit{Branch Purpose} – a reiteration of the overall project goal and the specific rationale for creating this branch (as defined at \texttt{BRANCH}); (2) \textit{Previous Progress Summary} – a coarse-grained summary of the branch's history, generated by giving the last commit's Previous Progress Summary and This Commit's Contribution; and (3) \textit{This Commit's Contribution} – a detailed narrative of what was achieved in the current commit. 

The second file is \textbf{\texttt{log.md}}, which stores the fine-grained reasoning trace of the agent’s execution. This includes every OTA (Observation–Thought–Action) cycle that occurs between commits. Each reasoning step is appended to \texttt{log.md} in real-time, forming a continuous trace of low-level decision-making. Upon committing, the relevant slice of this log is referenced to construct the summary in \texttt{commit.md}.

Finally, \textbf{\texttt{metadata.yaml}}, captures structured meta-level information. It includes details such as the current file structure of the project, per-file responsibilities, environment configurations, dependency graphs, or module interfaces. By default, commonly useful segments like \texttt{file\_structure} and \texttt{env\_config} are defined, while additional entries can be manually added by human users as needed. This file is updated on demand—typically during or after a \texttt{COMMIT}, when structural or configuration changes are detected.
Together, these files provide a layered, interpretable view of agent reasoning from abstract goals to step-level execution.

\subsection{GCC Commands}
The \texttt{Git-Context-Controller} exposes a set of agent-callable commands that allow reasoning models to manage, structure, and retrieve context in a durable and inspectable way. These commands include \texttt{COMMIT}, \texttt{BRANCH}, \texttt{MERGE}, and \texttt{CONTEXT}. 
These commands' function and usage are given to the agents in the system prompts, then the agents are encouraged to use them when needed. For example, when the agent reflects on its reasoning and detects a shift in direction, it would evaluate whether a \texttt{BRANCH} is warranted. When a reasoning subgoal is achieved, it is encouraged to call \texttt{COMMIT} and summarize the step. Below, we detail the purpose, usage, and implementation of each command.

\paragraph{\texttt{COMMIT <summary>}}
The \texttt{COMMIT} command is called when the agent identifies that its recent reasoning has resulted in a coherent and meaningful milestone, such as implementing a function, completing a test, or resolving a subgoal. Once invoked, the controller performs a structured update across multiple files within the current branch directory.

Specifically, let the agent’s recent execution trace be denoted as \(\mathcal{H}_t\). The \texttt{COMMIT} operation transforms this transient reasoning history into a persistent memory record
\[
\mathcal{M}_t = (I_t, S_t, D_t) = \texttt{COMMIT}(\mathcal{H}_t, S_{t-1}),
\]
where \(I_t\) represents the branch intent, \(S_t\) is a regenerated coarse-grained summary combining the previous commit summary \(S_{t-1}\) with the newly completed work, and \(D_t\) is a detailed description of the specific progress achieved since the last commit. When the global project plan evolves, \texttt{COMMIT} further induces an update of the global roadmap \texttt{main.md}. Finally, the memory update and code changes are consolidated into a versioned state using a Git commit with message \(S_t\). Through this transformation, a loose sequence of observation–thought–action steps is converted into a coherent and retrievable memory unit that supports long-horizon progress tracking and rollback.

\paragraph{\texttt{BRANCH <name>}}
The \texttt{BRANCH} command is called when the agent detects a meaningful divergence in direction, such as exploring an alternative algorithm, implementing a parallel module, or testing a new design hypothesis.

Let the current memory state be represented by the latest committed record \(\mathcal{M}_{t-1}\). When the command \texttt{BRANCH <name>} is issued, a new branch-specific execution state is created as:
$\mathcal{B}_t^{(name)} = \texttt{BRANCH}(\mathcal{M}_{t-1})$,
which initializes an empty observation–thought–action trace \(\mathcal{H}_t^{(name)}\) stored in \texttt{log.md}, together with a new structured memory file \texttt{commit.md} that records the intent and motivation of the branch. This transformation establishes an isolated reasoning trajectory in which alternative hypotheses or experimental workflows can be explored independently from the mainline history, while remaining fully inspectable and reversible through the same commit-based memory mechanism.

\paragraph{\texttt{MERGE <branch>}: }
The \texttt{MERGE} command is used when a branch has reached a conclusion and its results are ready to be integrated into the main plan. Before merging, the controller automatically calls \texttt{CONTEXT} on the target branch to surface its historical summaries and planning rationale.

Let the current branch memory state be \(\mathcal{M}_t = (I_t, S_t, D_t, \mathcal{H}_t)\) and the target branch memory state be \(\mathcal{M}_t^{(b)} = (I_t^{(b)}, S_t^{(b)}, D_t^{(b)}, \mathcal{H}_t^{(b)})\). When the command \texttt{MERGE} is issued, a new unified memory state is produced as
$
\mathcal{M}_{t+1} = \texttt{MERGE}(\mathcal{M}_t, \mathcal{M}_t^{(b)}),
$
where the updated memory record stored in \texttt{commit.md} is defined by
\[
(S_{t+1}, D_{t+1}) = \mathcal{F}_{\text{merge}}\big((S_t, D_t), (S_t^{(b)}, D_t^{(b)})\big).
\]

Here, \(\mathcal{F}_{\text{merge}}(\cdot)\) denotes a synthesis operator that integrates the branch purpose and progress from both branches into a unified summary \(S_{t+1}\) and a detailed description \(D_{t+1}\) explaining the rationale and outcome of the merge. The global planning state \texttt{main.md} is simultaneously updated as
$
\texttt{main.md}_{t+1} = \mathcal{G}(\texttt{main.md}_t, \mathcal{M}_t^{(b)}),
$
reflecting the impact of the merged branch on the overall roadmap and future milestones.

The execution traces stored in \texttt{log.md} are combined as
$
\mathcal{H}_{t+1} = \mathcal{H}_t \cup \mathcal{H}_t^{(b)},
$
with explicit origin annotations to preserve the provenance of observation–thought–action steps. Finally, the unified memory state \(\mathcal{M}_{t+1}\) is checkpointed through a new Git commit, ensuring that symbolic memory and executable artifacts are synchronized and versioned.

\paragraph{\texttt{CONTEXT <options>}: }
The \texttt{CONTEXT} command allows agents to retrieve memory at multiple levels of granularity, from global overviews to fine-grained token-level execution traces. This supports both reflective reasoning and task continuation across sessions. Agents are required to call \texttt{CONTEXT} in specific scenarios, such as when a new agent resumes an ongoing task, or before the MERGE command. Besides that, the agent is able to call \texttt{CONTEXT} proactively whenever it finds context retrieval necessary.

When the agent issues the \texttt{CONTEXT} command, the controller returns a structured snapshot of the current project state, analogous to a \texttt{git status} view over the memory directory. This snapshot exposes the global project purpose and milestone progress derived from \texttt{main.md}, together with the set of available branches.

For branch-level inspection, the agent may request \texttt{CONTEXT --branch <branch>}, which retrieves the branch intent and the latest progress summary stored in \texttt{commit.md}, along with a bounded window of recent commit records. Let the ordered commit history of a branch be denoted as \(\{\mathcal{M}_i\}_{i=1}^T\). The returned view corresponds to a windowed projection
$
\mathcal{V}_k = \{\mathcal{M}_i\}_{i=k}^{k+K},
$
where \(K\) is a fixed context budget and \(k\) is controlled by scrolling operations. This design ensures that long reasoning histories can be traversed incrementally without exceeding the agent’s context capacity.
More fine-grained retrieval is supported through specialized queries. The command \texttt{CONTEXT --commit <hash>} returns the complete structured memory record associated with a specific commit, while \texttt{CONTEXT --log} exposes a windowed segment of the execution trace \(\mathcal{H}\) recorded in \texttt{log.md} using the same sliding-window mechanism.
System-level metadata, such as file structure and environment configuration, is accessed through \texttt{CONTEXT --metadata <segment>}, which retrieves the corresponding portion of \texttt{metadata.yaml}.

\section{Experiment}

\begin{table*}[t!]
\centering\small
\setlength{\tabcolsep}{7pt}
\resizebox{0.95\textwidth}{!}{
\begin{tabular}{lll|cc|cc}
\toprule
& & & \multicolumn{2}{c|}{\textbf{BrowseComp-Plus}} & \multicolumn{2}{c}{\textbf{SWE-Bench Verified}} \\
\textbf{Model} & \textbf{Peak Length} & \textbf{Max \#Token} & \textbf{Pass\@1} & \textbf{Tool Calls} & \textbf{Pass\@1} & \textbf{Tool Calls} \\
\midrule

\rowcolor{gray!10}\multicolumn{7}{c}{\textbf{ReAct Agent}} \\
GPT-5   & 327K & 327K & 0.793 & 14.2 & 0.718   & 42.6 \\
GPT-4.1  & 327K & 327K & 0.640 & 5.6 & 0.486   & 28.7 \\
DeepSeek-V3.1  & 327K & 327K & 0.613 & 10.6 & 0.610   & 53.2 \\
GLM-4.5-Air  & 327K & 327K   & 0.566    & 11.1 & 0.576   & 51.2 \\
Qwen3-235B-A22B  & 327K & 327K & 0.560 & 12.8 & 0.344   & 32.1 \\
Claude 4 Sonnet        & 327K & 327K & 0.672 & 13.6 & 0.682   & 48.3 \\
\midrule

\rowcolor{gray!10}\multicolumn{7}{c}{\textbf{SWE Agent}} \\

GPT-5              & 327K & 327K & - & - & 0.650 & 70.3 \\
GPT-4.1            & 327K & 327K & - & - & 0.396 & 68.5 \\
DeepSeek-V3.1      & 327K & 327K & - & - & 0.420 & 72.2 \\
GLM-4.5-Air        & 327K & 327K & - & - & 0.542 & 71.7 \\
Qwen3-235B-A22B    & 327K & 327K & - & - & 0.406 & 60.6 \\
Claude 4 Sonnet    & 327K & 327K & - & - & 0.666 & 62.4 \\
\midrule
\rowcolor{gray!10}\multicolumn{7}{c}{\textbf{Summary Agent}} \\

GPT-5              & 327K & 327K $\times$ 100 & 0.765 & 16.4 & 0.690 & 52.0 \\
GPT-4.1            & 327K & 327K $\times$ 100 & 0.633 & 12.3 & 0.474 & 49.3 \\
DeepSeek-V3.1      & 327K & 327K $\times$ 100& 0.592 & 18.6 & 0.626 & 55.6 \\ 
GLM-4.5-Air        & 327K & 327K $\times$ 100 & 0.565 & 14.3 & 0.566 & 51.0 \\
Qwen3-235B-A22B    & 327K & 327K $\times$ 100 & 0.578 & 10.4 & 0.378 & 44.6 \\
Claude 4 Sonnet    & 327K & 327K $\times$ 100 & 0.685 & 11.8 & 0.666 & 47.1 \\
\midrule
\rowcolor{gray!10}\multicolumn{7}{c}{\textbf{Folding Agent}} \\

GPT-5              & 327K & 327K $\times$ 100 & 0.815 & 20.1 & 0.746 & 95.3 \\
GPT-4.1            & 327K & 327K $\times$ 100 & 0.665 & 16.3 & 0.626 & 88.4 \\
DeepSeek-V3.1      & 327K & 327K $\times$ 100 & 0.640 & 18.2 & 0.616 & 96.6 \\
GLM-4.5-Air        & 327K & 327K $\times$ 100 & 0.595 & 19.7 & 0.596 & 92.9 \\
Qwen3-235B-A22B    & 327K & 327K $\times$ 100 & 0.585 & 20.3 & 0.366 & 82.7 \\
Claude 4 Sonnet    & 327K & 327K $\times$ 100 & 0.720 & 22.6 & 0.740 & 84.1 \\
\midrule
\rowcolor{ForestGreen!20}\multicolumn{7}{c}{\textbf{GCC Agent}} \\

GPT-5              & 327K & 327K $\times$ 100 & \textbf{0.834} {\color{ForestGreen}(+1.9)} & 24.5 & \textbf{0.790} {\color{ForestGreen}(+4.5)} & 101.5 \\
GPT-4.1            & 327K & 327K $\times$ 100& \textbf{0.722} {\color{ForestGreen}(+5.7)} & 21.0 & \textbf{0.644} {\color{ForestGreen}(+1.9)} & 98.7 \\
DeepSeek-V3.1      & 327K & 327K $\times$ 100 & \textbf{0.693} {\color{ForestGreen}(+5.3)} & 19.2 & \textbf{0.662} {\color{ForestGreen}(+4.6)} & 118.2 \\
GLM-4.5-Air        & 327K & 327K $\times$ 100 & \textbf{0.655} {\color{ForestGreen}(+6.0)} & 26.6 & \textbf{0.634} {\color{ForestGreen}(+3.9)} & 112.5 \\
Qwen3-235B-A22B    & 327K & 327K $\times$ 100 & \textbf{0.621} {\color{ForestGreen}(+3.6)} & 21.4 & \textbf{0.478} {\color{ForestGreen}(+6.3)} & 95.4 \\
Claude 4 Sonnet    & 327K & 327K $\times$ 100 & \textbf{0.786} {\color{ForestGreen}(+6.6)} & 28.8 & \textbf{0.802} {\color{ForestGreen}(+6.2)} & 110.7 \\

\bottomrule

\end{tabular}}
\caption{\textbf{Performance on BrowseComp-Plus (N=150) and SWE-Bench Verified (N=500).}}\label{tab:main}
\vspace{-20pt}
\end{table*}

\subsection{Datasets}
\textbf{Deep Research: BrowseComp-Plus.}
For research-oriented tasks, we rely on BrowseComp-Plus (BC-Plus)~\citep{Chen2025BrowseCompPlusAM}, an extension of BrowseComp enriched with verified targets. 
High-quality supervision is particularly important in this domain, yet most existing web-research datasets are not publicly released~\citep{Qiao2025WebResearcherUU,Li2025WebSailorV2BT}. 
We evaluate GCC on all instances of the dataset. 
The agent interacts with two tools, \texttt{search(query, topk)} and \texttt{open\_page(url)}.
All retrieval is performed with Qwen3-Embed-8B.

\textbf{Agentic SWE: SWE-Bench and SWE-Benchlite.}
For software engineering, we follow the widely used SWE-Bench benchmark~\citep{swebench}, which tests an agent’s ability to produce correct patches for real-world bugs described in natural-language issue reports. 
A task consists of the buggy project snapshot and the patch generation requirement. 
In addition, we also report results on SWE-Benchlite~\citep{swebenchlite}, a curated and self-contained subset of 300 higher-quality problems that has become standard in recent evaluations.

\subsection{Implementation}
We set the LLM context window to 32,768 tokens for all comparison models. For the fold agent, summary agent, and GCC, we allow up to 100 active summarized archives/branches, yielding a theoretical maximum accessible context of 3,276,800 tokens. In all experiments, the commit retrieval window of the \texttt{CONTEXT} operation is fixed to \(K=1\), such that only the most recent commit record is revealed to the agent. 
When evaluated on SWE-Bench, GCC is integrated on top of the standard SWE-Agent framework.

\subsection{Baselines}
We compare \tech against four standard long-context agent baselines widely adopted in prior work.  
Each baseline is instantiated with the same set of backend models as in Table~\ref{tab:main}:  
GPT-5, GPT-4.1, DeepSeek-V3.1, GLM-4.5-Air, Qwen3-235B-A22B, and Claude 4 Sonnet.  
All agents operate under identical tool APIs, context budgets, and evaluation settings.
\noindent \textbf{ReAct Agent}~\citep{Yao2022ReActSR}.  
Maintains full interaction history with a ReAct-style reasoning–acting loop.  
We evaluate under a fixed 327K token budget (Peak Length and Max\#Token).
\noindent \textbf{SWE-Agent}~\citep{Jimenez2023SWEbenchCL}.  
The default open-source reference system for SWE-Bench.  
Iteratively generates patches using execution, editing, and testing tool calls while keeping uncompressed full context.  
Following prior work, SWE-Agent is evaluated on SWE-Bench Verified only.
\noindent \textbf{Summary Agent}~\citep{Yu2025MemAgentRL}.  
Uses summary-based context compression: when the context approaches the 327K limit, it produces a high-level summary and replaces earlier history.  
This represents the standard “summarize-when-full’’ policy.
\noindent \textbf{Folding Agent}.  
Implements the Folding paradigm \cite{sun2025scaling}, periodically compressing earlier reasoning into structured summaries (“folds’’) while maintaining a shorter active window.  
This provides a middle ground between pure retention and pure summarization.
\noindent All four baselines use the same underlying models, data, tools, and evaluation protocol as \tech.

\section{Experimental Results}
Table~\ref{tab:main} presents a comprehensive comparison across four long-context agent baselines, ReAct, SWE-Agent, Summary Agent, and Folding Agent—together with our proposed GCC agent, evaluated under identical model backends and tool APIs. Several consistent patterns emerge.

\paragraph{GCC surpasses all controlled baselines.}
Across all model backbones, GCC achieves the best Pass@1 on both BrowseComp-Plus and SWE-Bench Verified. For instance, on BrowseComp-Plus with GPT-5, GCC reaches a Pass@1 of 0.834, outperforming the next-best Folding Agent (0.815). This trend persists across smaller models: with GPT-4.1, GCC improves Pass@1 from 0.665 (Folding) to 0.722; with DeepSeek-V3.1, from 0.640 to 0.693; and similarly strong gains appear for GLM-4.5-Air, Qwen-235B-A22B, and Claude 4 Sonnet. These results demonstrate that GCC consistently enhances the underlying model’s capability, regardless of backend architecture.

\paragraph{GCC yields even larger gains on SWE-Bench Verified.}
SWE-Bench Verified presents significantly greater complexity, yet GCC achieves the strongest performance across all backbones. Using GPT-5, GCC attains a Pass@1 of 0.790, exceeding the Folding Agent (0.746) and outperforming the Summary and ReAct agents by even larger margins. Similar improvements appear across other models: for example, GCC improves Claude 4 Sonnet from 0.740 to 0.802, and DeepSeek-V3.1 from 0.616 to 0.662. These uniform improvements on a challenging benchmark confirm GCC’s advantage in handling long-horizon software reasoning.

\begin{table*}[ht]
\centering
\rowcolors{3}{white}{gray!10} 
\scalebox{0.90}{
\begin{tabular}{ll|rrrrrr}
\toprule
 \multirow{2}*{Tool} & \multirow{2}*{\llm} & \multirow{2}*{\%{} Resolved} & \multirow{2}*{\makecell{Avg.\\{}\$ Cost}} & \multirow{2}*{\makecell{Avg.\\{}\# Tokens}} & \multicolumn{3}{c}{\% Correct Location} \\
 & & & & & Line & Function & File \\
\midrule

\codestoryaide{}~\cite{codestoryaide}  & \makecell[l]{ \gptfouro{}+ \claudesonnet} & 129 (43.00\%) &  - & - &41.7\% &58.7\% &72.0\% \\
\bytedanceagent{}~\cite{marscode}  & NA & 118 (39.33\%) &  - & - &42.7\% &58.0\% &79.7\% \\
\honeycomb{}~\cite{honeycomb}  & NA & 115 (38.33\%) &  - & - &44.3\% &57.0\% &69.3\% \\
\mentatbot{}~\cite{mentatbot}  &  \gptfouro{} & 114 (38.00\%) &  - & - &37.3\% &53.3\% &69.3\% \\
\gru{}~\cite{gru}  & NA & 107 (35.67\%) &  - & - &38.3\% &54.3\% &75.0\% \\
\isoform{}~\cite{isoform}  & NA & 105 (35.00\%) &  - & 41,963 &38.7\% &55.3\% &72.0\% \\
\supercoder{}~\cite{supercoder}  & NA & 102 (34.00\%) &  - & - &41.7\% &63.7\% &65.7\% \\
\lingma{}~\cite{lingma}  & \makecell[l]{ \gptfouro{}+ \claudesonnet} & 99 (33.00\%) &  - & - &40.0\% &58.7\% &75.0\% \\
\factorydroid{}~\cite{factorydroid}  & NA & 94 (31.33\%) &  - & - &36.7\% &55.7\% &72.7\% \\
\amazonqagent{}-v2~\cite{amazonqdeveloper}  & NA & 89 (29.67\%) &  - & - &40.3\% &52.0\% &74.3\% \\
\specrover{}~\cite{specrover}  & \makecell[l]{ \gptfouro{}+ \claudesonnet} & 93 (31.00\%) & \$0.65 & - & - & - & - \\
\coder{}~\cite{coder}  &  \gptfour{} & 85 (28.33\%) & \$3.34 & 323,802 &35.7\% &52.3\% &67.0\% \\
\masai{}~\cite{masai}  & NA & 84 (28.00\%) &  - & - &38.7\% &56.3\% &75.0\% \\
\sima{}~\cite{sima}  &  \gptfouro{} & 83 (27.67\%) & \$0.82 & - &37.0\% &54.0\% &79.0\% \\
\ibmagent{}~\cite{ibmagent}  & NA & 80 (26.67\%) &  - & - &39.7\% &56.7\% &73.3\% \\
\opencsgagent{}~\cite{opencsgstarship}  &  \gptfour{} & 71 (23.67\%) &  - & - &39.0\% &61.7\% &90.7\% \\
\amazonqagent{}~\cite{amazonqdeveloper}  & NA & 61 (20.33\%) &  - & - &34.0\% &43.7\% &71.7\% \\
\repounderstander{}~\cite{repounderstander}  &  \gptfour{} & 64 (21.33\%) &  - & - & - & - & - \\
\autocoderover{}-v2~\cite{autocoderovertwo} &  \gptfouro{} & 92 (30.67\%) &  - & - &35.0\% &52.3\% &69.3\% \\
\repograph~\cite{repograph} &  \gptfouro{} & 89 (29.67\%) &  - & - &36.7\% &51.3\% &71.0\% \\
\multirow{1}*{\moatless{}~\cite{moatless}} &  \claudesonnet{} & 80 (26.67\%) & \$0.17 & - &38.7\% &54.7\% &78.7\% \\
 &  \gptfouro{} & 74 (24.67\%) & \$0.14 & - &36.0\% &52.0\% &73.0\% \\
\opendevincodeact{}~\cite{opendevin} &  \claudesonnet & 80 (26.67\%) & \$1.14 & - &38.0\% &49.7\% &67.3\% \\
\aider{}~\cite{aidar} & \makecell[l]{ \gptfouro{}+ \claudesonnet} & 79 (26.33\%) &  - & - &35.3\% &50.0\% &69.7\% \\
\multirow{1}*{\sweagent{}~\cite{sweagent}} &  \claudesonnet{} & 69 (23.00\%) & \$1.62 & 521,208 &40.7\% &54.3\% &72.0\% \\
 &  \gptfouro{} & 55 (18.33\%) & \$2.53 & 498,346 &29.3\% &42.3\% &58.3\% \\
 &  \gptfour{} & 54 (18.00\%) & \$2.51 & 245,008 &30.7\% &45.3\% &61.0\% \\
\appmapnavie{}~\cite{appmapnavie} &  \gptfouro{} & 65 (21.67\%) &  - & - &29.7\% &44.7\% &59.7\% \\
\autocoderover{}~\cite{autocoderover} &  \gptfour{} & 57 (19.00\%) & \$0.45 & 38,663 &29.0\% &42.3\% &62.3\% \\
\rag{}~\cite{sweagent} &  \claudeopus{} & 13 (4.33\%) & \$0.25 & - &22.0\% &30.0\% &57.0\% \\
 &  \gptfour{} & 8 (2.67\%) & \$0.13 & - &12.7\% &23.3\% &47.3\% \\
 &  \claude{}-3 Opus & 9 (3.00\%) &  - & - &16.7\% &24.3\% &46.7\% \\
 &  \chatgpt{} & 1 (0.33\%) &  - & - &6.3\% &11.3\% &27.3\% \\
AgentLess~\cite{xia2024agentless} &  \gptfouro{} & 96 (32.00\%) & \$0.70 & 78,166 &35.3\% &52.0\% &69.7\% \\
\midrule
\textbf{GCC} &  \chatgpt{} & \textbf{90 (30.00\%)} & \$0.57 & 386,490 &35.7\% &51.3\% &69.3\% \\
& \gptfouro{} & \textbf{138 (46.00\%)} & \$1.13 & 546,798       & 42.0\% & 59.3\% & 76.0\% \\
& \claudesonnet{} & \textbf{144 (48.00\%)} & \$1.21 & 468,549 &44.3\% &61.7\% &78.7\% \\

\bottomrule

\end{tabular}
}
\caption{Results on SWEBench-Lite. 
The ‘–’ symbol denotes missing / unreleased information needed to compute the corresponding value.
\claudesonnet{} denotes \claudesonnet{}onnet.}\label{tab:code}
\vspace{-20pt}
\end{table*}

\paragraph{GCC allocates substantially more computation and interaction.}
A distinguishing characteristic of GCC is its increased tool-call frequency. On BrowseComp-Plus with GPT-5, GCC uses 24.5 tool calls, compared to 20.1 for the Folding Agent and just 14.2 for the long-context ReAct baseline. This trend becomes even more pronounced on SWE-Bench Verified, where GCC increases tool calls from 95.3 (Folding Agent) to 101.5. Similar jumps occur across all model backbones. These results indicate that GCC not only compresses and manages context better but also learns to allocate interaction capacity adaptively—probing deeper, exploring more paths, and conducting more thorough reasoning during thinking.

\subsection{GCC Performance-Efficiency on Coding}\label{sec:close}
We report an extended analysis of the SWE-Benchlite results in Table~\ref{tab:code}, comparing \tech{} against 26 state-of-the-art agentic coding systems spanning open-source methods, commercial closed-source agents, hybrid GPT–Claude ensembles, and retrieval-only baselines. Because these systems differ widely in model backend, memory strategy, and interaction protocol, SWE-Benchlite provides a stress test of *general-purpose coding competence* rather than controlled ablations. Several fine-grained patterns emerge.

\paragraph{GCC establishes a new state of the art across all competing systems.}
\tech{} achieves a resolution rate of \textbf{48.00\%}, the highest among all 26 systems. The next best performer, CodeStory Aide (43.00\%), uses a hybrid GPT-4o + Claude 3.5 Sonnet backend but does not release intermediate reasoning or memory structure. Other top-tier commercial agents: ByteDance MarsCode (39.33\%), Honeycomb (38.33\%), and MentatBot (38.00\%), all fall short by 5–10 percentage points. This margin is remarkable because many of these competing systems rely on fine-tuned internal components or private coding infrastructure, whereas \tech{} uses a transparent protocol layered onto an off-the-shelf LLM.

\begin{figure*}[t]
\centering
\includegraphics[width=0.95\linewidth]{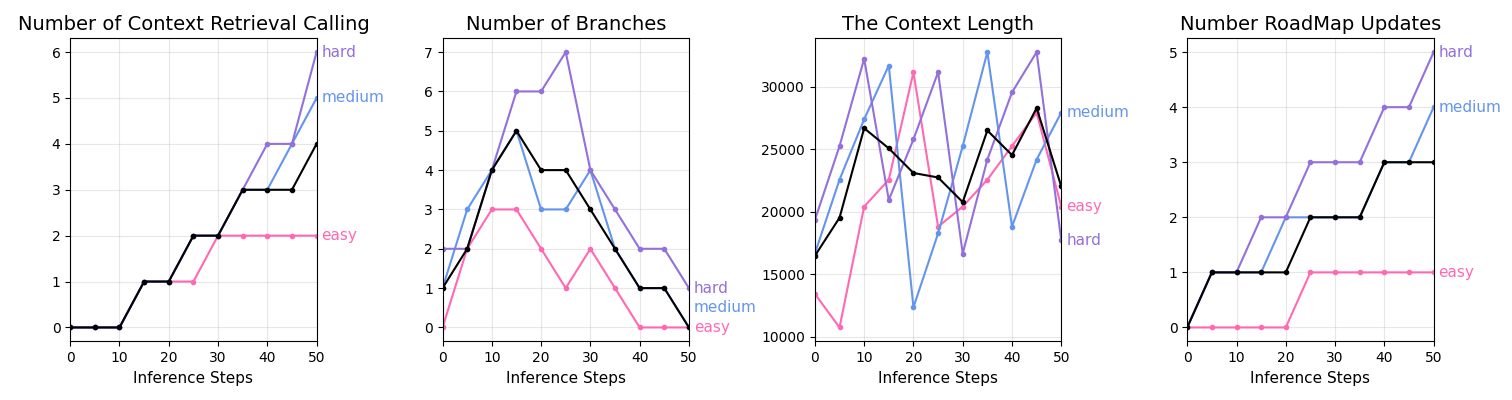}
\vspace{-5pt}
\caption{Number of context retrieval calling, number of branches, the context length and number of RoadMap updates with the increase of the inference steps}
\label{fig:performance}
\vspace{-10pt}
\end{figure*}

\paragraph{GCC closes the gap between proprietary agents and model-agnostic open systems.}
A notable trend in Table~\ref{tab:code} is that several commercial systems (e.g., MarsCode, SIMA, Lingma, Honeycomb) cluster within a narrow accuracy band of 33–40\%. Their performance gains primarily come from heavy engineering, test harness integration, and custom model routing: advantages unavailable to typical open-source agents. In contrast, \tech{} surpasses all of them while remaining purely protocol-driven. This suggests that structured, versioned memory confers benefits comparable to (or exceeding) architecture-level fine-tuning and proprietary engineering.

\paragraph{GCC achieves high performance without excessive inference cost.}
Despite the increased structure and deeper exploration that GCC encourages, the cost per task remains modest. For example, on Claude 3.5 Sonnet, the average cost is \$1.21—comparable to, and often lower than, many top-performing agents such as SWE-Agent (up to \$2.53) or CodeR (up to \$3.34). This shows that the performance gain is not merely due to brute-force trial-and-error or excessive tool usage; rather, GCC induces structured tool usage that is cost-efficient relative to the quality of patches produced.

\subsection{Ablation and Analysis}\label{sec:ablation}

We perform an ablation study to evaluate the contribution of each component in GCC, including the RoadMap (main.md), Detailed Logs (log.md), Meta Data (metadata.yaml), CONTEXT retrieval, and BRANCH \& MERGE, and COMMIT operations. Results on SWE-Bench Verified are reported in Table~\ref{tab:comparison}.

Starting from a raw Claude-4-Sonnet without structured memory (67.2\%), adding RoadMap and COMMIT improves performance to 69.1\%, showing that milestone-based checkpointing alone provides limited gains. Introducing Detailed Logs and CONTEXT retrieval yields a substantial improvement to 75.3\%, highlighting the importance of preserving fine-grained reasoning traces and enabling explicit historical access. Further incorporating Meta Data increases performance to 77.8\%, suggesting that structured architectural context supports consistency in long-horizon reasoning. The full system with BRANCH and MERGE achieves the best result of 80.2\%, demonstrating that isolated exploration and controlled synthesis of alternative reasoning paths are critical for complex tasks.

We further analyze the evolution of internal behaviors during inference in Figure~\ref{fig:performance}, including the number of CONTEXT calls, active branches, context length, and RoadMap updates.
To enable controlled comparison across levels of challenge, we group evaluation tasks into \emph{easy}, \emph{medium}, and \emph{hard}.  
For SWE datasets, difficulty is derived from the human time-to-resolution metadata originally provided with SWE-Bench: issues historically fixed within 15 minutes are \emph{easy} (194 tasks), those requiring 15--60 minutes are \emph{medium} (261 tasks), and those taking more than an hour are \emph{hard} (45 tasks).

We find in Figure~\ref{fig:performance}, CONTEXT retrieval increases with inference steps and task difficulty, indicating adaptive reliance on historical memory. The number of branches rises in intermediate stages and decreases as the agent converges, reflecting exploration followed by consolidation. Context length shows non-monotonic variation due to periodic summarization and retrieval, preventing unbounded growth. RoadMap updates increase steadily, especially for harder tasks, revealing continuous refinement of the global plan. These results show that each GCC component contributes incrementally, while the complete system induces structured behaviors of exploration, reflection, and consolidation, leading to more effective long-horizon reasoning.

\begin{table}[h]
    \centering
    \resizebox{0.95\hsize}{!}{
     \begin{tabular}{ccc|ccc|c}
        \hline
        RoadMap& Detailed  & Meta  & CONTEXT & BRANCH \&  & COMMIT & \textbf{SWE} \\
        & Logs & Data &  & MERGE &  & Verified \\
        \hline
         &  &  &  &  &  & 67.2 \\
        \checkmark &  &  &  &  & \checkmark  & 69.1 \\
        \checkmark & \checkmark &  & \checkmark &  & \checkmark & 75.3 \\
        \checkmark & \checkmark & \checkmark  & \checkmark &  & \checkmark  & 77.8 \\ \hline
        \checkmark & \checkmark & \checkmark & \checkmark & \checkmark & \checkmark & 80.2 \\ 
        \hline
    \end{tabular}}
    \caption{Ablation study on GCC File System and Commands.}
    \vspace{-10pt}
    \label{tab:comparison}
\end{table}

\section{Conclusion}
We introduced GCC, a structured context management framework that organizes agent memory using version control-inspired operations. GCC enables agents to persist, retrieve, and explore reasoning trajectories through committing, branching, and merging.
Experiments show that GCC yields consistent performance gains and achieves SOTA results, suggesting that structured, Git-style memory is a key ingredient for effective long-horizon reasoning in autonomous agents.
\clearpage


\bibliography{example_paper}
\bibliographystyle{icml2026}

\end{document}